# Thermoelectric transport of massive Dirac fermions in bilayer graphene

Seung-Geol Nam,[1] Dong-Keun Ki,[1,2] and Hu-Jong Lee[1,*]

[1]*Department of Physics, Pohang University of Science and Technology, Pohang, Republic of Korea*
[2]*DPMC and GAP, University of Geneva, 24 Quai Ernest-Ansermet, CH1211 Geneva, Switzerland*

Thermoelectric power (TEP) is measured in bilayer graphene for various temperatures and charge-carrier densities. At low temperatures, measured TEP well follows the semiclassical Mott formula with a hyperbolic dispersion relation. TEP for a high carrier density shows a linear temperature dependence, which demonstrates a weak electron-phonon interaction in the bilayer graphene. For a low carrier density, a deviation from the Mott relation is observed at high temperatures and is attributed to the low Fermi temperature in the bilayer graphene. Oscillating TEP and the Nernst effect for varying carrier density, observed in a high magnetic field, are qualitatively explained by the two dimensionality of the system.

## I. INTRODUCTION

Recently, gate voltage ($V_{BG}$) and temperature ($T$) dependences of the thermoelectric power (TEP) were measured in monolayer graphene, a single atomic layer of graphite.[1–3] TEP is defined as the ratio of the thermoelectric voltage ($V_{th}$) to the temperature difference ($\Delta T$) across a sample. Since TEP is the electrical response to the thermodynamic perturbation, it is sensitive to the electronic structure near the Fermi energy. TEP in monolayer graphene was found to be dominated by diffusive carriers in the temperature range of $T<300$ K. Thus, by employing the semiclassical Mott relation,[4] the linear-dispersion relation as well as the nature of impurity scattering in monolayer graphene has been investigated.[1,2,5]

Bilayer graphene, a stack of two weekly coupled graphene layers, also has attracted high interest owing to its plausible device applications[6,7] as well as fundamental interests.[8] Recently, the transverse thermoelectric conductivity ($\alpha_{xy}$) of bilayer graphene was measured, mainly in the quantum Hall regime.[9] In the report, $\alpha_{xy}$ increases linearly at low $T$ and saturates close to the predicted universal value at high $T$. Also, an anomaly in $\alpha_{xy}$ is found near the charge neutrality point (CNP), which may be related to the counterpropagating edge states at the CNP. Compared with the case of monolayer graphene, however, reports on TEP measurements in bilayer graphene are still scarce. Similar to monolayer graphene, TEP measurements in bilayer graphene in zero magnetic field are also expected to provide the detailed information on the dispersion relation, the electron-phonon interaction,[10] and the yet controversial scattering mechanism in the system.[11–14] In addition, the transverse component of TEP (the Nernst signal) in the quantum Hall regime can be used to examine the dimensionality of graphene stacks.[15–17]

In this study, we report the gate-voltage dependence of TEP taken in bilayer graphene for different temperatures and magnetic fields. By assuming hyperbolic dispersion relation, we confirm that measured TEP shows a good agreement with the Mott relation at low temperatures, which demonstrates the validity of the semiclassical Boltzmann theory in this system. The phonon-drag effect was not observed, indicating a weak electron-phonon coupling in bilayer graphene. For a low carrier density, however, distinct from the monolayer graphene, a deviation from the Mott relation is observed along with a saturating tendency of the TEP at high temperatures, which we attribute to the low Fermi temperature ($T_F$) of the bilayer graphene. In a high magnetic field, TEP and the Nernst signal showed oscillations with changing back-gate voltage, which are qualitatively explained by the two dimensionality of the system.

## II. SAMPLE PREPARATION AND MEASUREMENTS

The inset of Fig. 1(b) shows the typical configuration of devices. Bilayer graphene was prepared by the mechanical exfoliation on 300 nm SiO$_2$/highly doped Si substrate. After a piece of suitable bilayer graphene is selected under microscope, electrodes were attached by e-beam patterning and subsequent Ti/Au (3/17 nm) evaporation. Details of the fabrication processes are reported elsewhere.[18] After the sample fabrication, quantum Hall effect was measured to confirm the number of graphene layers in a sample. Typical mobility of our sample was 2000–4000 cm$^2$/V s. To set up a temperature difference $\Delta T$ ($\ll T$) across a sample, a sufficient current was applied through a meandering heater line and $\Delta T$ was determined from the four-terminal resistances of two local thermometers shown in the inset of Fig. 1(b). The thermoelectric voltage $V_{th}$ was measured by the ac lock-in technique, where the second-harmonic signal was detected as a low-frequency ac current ($f<2$ Hz) was driven to the microheater. Similar technique of setting up $\Delta T$ and measuring $V_{th}$ were employed in earlier studies.[1–3,19] The TEP ($S_{xx}$) and the Nernst coefficient ($S_{yx}$) are obtained from the relations $S_{xx}=-E_x/|\nabla T|$ and $S_{yx}=E_y/|\nabla T|$, respectively. Here, $E_{x(y)}$ is the longitudinal (transverse) component of electric field and $\nabla T$ is the longitudinal temperature gradient. Measurements were made using a homemade cryostat in a temperature range of 4.2–300 K and in magnetic fields up to 7 T.

## III. RESULTS AND DISCUSSION

Zero-field two-terminal resistance $R$ and the longitudinal component of TEP $S_{xx}$ are shown as a function of the back-gate voltage $V_{BG}$ in Figs. 1(a) and 1(b), respectively. $R$ re-veals a maximum at the CNP ($V_{CNP} \sim -4$ V for this sample). Consistent with previous studies,[20,21] $R$ near the CNP increases with decreasing $T$ due to the thermal excitation of carriers, which is dominant near the CNP. Different from previous report,[20] however, a small but finite $T$ dependence of the resistance is observed away from the CNP, presumably due to the presence of electron-hole puddles.[22,23] As in monolayer graphene,[1–3] the ambipolar nature of the carriers manifests itself as the sign change in $S_{xx}$ in Fig. 1(b) at the CNP.

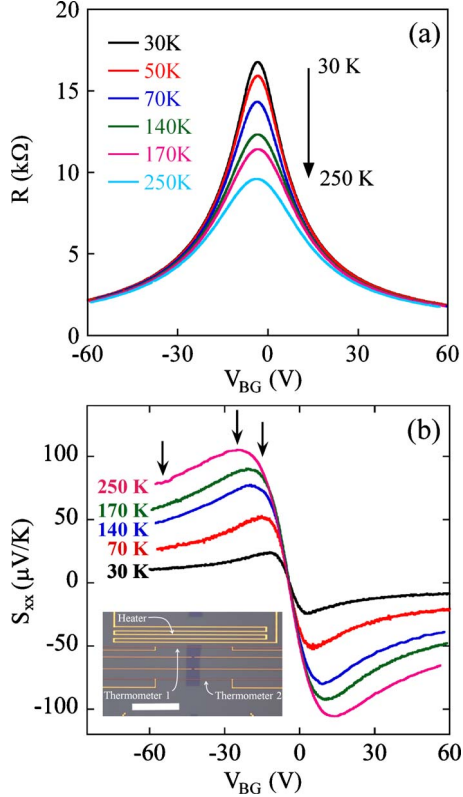

FIG. 1. (Color online) (a) Two-terminal resistance $R$ and (b) the TEP as a function of the backgate voltage $V_{BG}$ in zero magnetic field and at various temperatures: $T=30, 50, 70, 140, 170,$ and 250 K. Solid arrows indicate $V_{BG}$'s where data points in Figs. 3(a)–3(c) are taken. Inset: optical image of a typical device. Size of the scale bar is 30 $\mu$m.

For a quantitative analysis, we adopt the semiclassical Mott relation for TEP,[4]

$$S^{Mott} = -\frac{\pi^2 k_B^2 T}{3e} \frac{1}{G} \frac{dG}{dV_{BG}} \frac{dV_{BG}}{dE}\bigg|_{E=E_F}, \quad (1)$$

where $k_B$, $e$, and $G$ are the Boltzmann constant, the electron charge, and the conductance, respectively. The only experimentally undetermined term in Eq. (1) is $dV_{BG}/dE$ or the dispersion relation of the system $E(k)$, where the Fermi wave vector $k_F$ is related to $V_{BG}$ as $k_F = \sqrt{\pi|n_{e,h}|} = \sqrt{\pi\alpha|V_{BG}|}$ ($n_{e,h}$ is the electron or hole carrier density and $\alpha$ is a geometry-dependent capacitance of the device). According to a tight-binding model in first-order approximation,[24] the dispersion relation of bilayer graphene is hyperbolic as

$$E = \pm \frac{1}{2}\gamma_1[\sqrt{1 + 4v_F^2\hbar^2 k^2/\gamma_1^2} - 1], \quad (2)$$

where $\gamma_1$ is the interlayer hopping energy, $v_F$ is the Fermi velocity, and $\hbar$ is the Planck's constant.

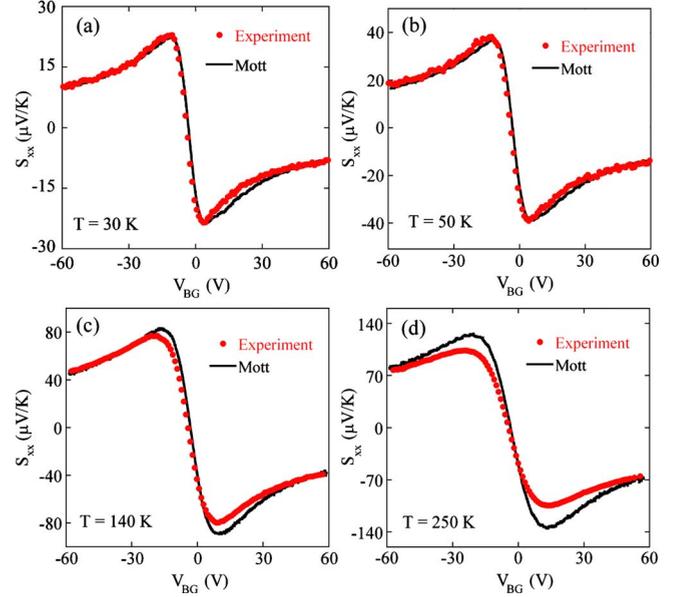

FIG. 2. (Color online) The TEP measured (filled circle) and calculated from the Mott relation (solid line) as a function of $V_{BG}$ at (a) $T=30$ K, (b) $T=50$ K, (c) $T=140$ K, and (d) $T=250$ K.

In Fig. 2, for various temperatures, we compare the measured TEP, $S_{xx}^{meas}$, with the ones estimated from the Mott relation, $S_{xx}^{Mott}$, by adopting the dispersion relation in Eq. (2). Here and afterward, we fix the fitting parameters as $v_F \cong 0.95 \times 10^6$ m/s and $\gamma_1 = 0.39$ eV, which are comparable to the results of earlier spectroscopic measurements.[25] At low temperatures of $T=30$ K and $T=50$ K [Figs. 2(a) and 2(b)], $S_{xx}^{meas}$ is in excellent agreement with $S_{xx}^{Mott}$, which demonstrates the validity of the semiclassical Boltzmann theory in this system.

Here, we note that, even in the presence of a perpendicular electric field that may induce a band gap at the CNP,[6,26] the gapless hyperbolic dispersion relation reproduces the experimental data very well. Moreover, a large enhancement of TEP,[27] which is expected in bilayer graphene with a large band gap, was not observed. For a high carrier density, our observation is understandable as the change in the density of states (DOS) at the Fermi energy due to the band-gap opening can be neglected.[26,12] However, the signature of the band-gap opening is also absent near the CNP where the change in the DOS is no longer negligible. We suspect that it was caused by large fluctuations of the disorder potential[23] compared to the size of the band gap.

At high temperatures, $S_{xx}^{meas}$ reveals a deviation from the $S_{xx}^{Mott}$ near the CNP and the deviation increases with $T$ [Figs. 2(c) and 2(d)]. Besides, with increasing $T$, the range of $V_{BG}$ where the deviation occurs becomes wider. For detailed analysis, we plot $S_{xx}^{meas}$ and $S_{xx}^{Mott}$ as a function of $T$ in Figs. 3(a)–3(c) for different fixed values of $V_{BG}$'s marked by solid arrows in Fig. 1(b).

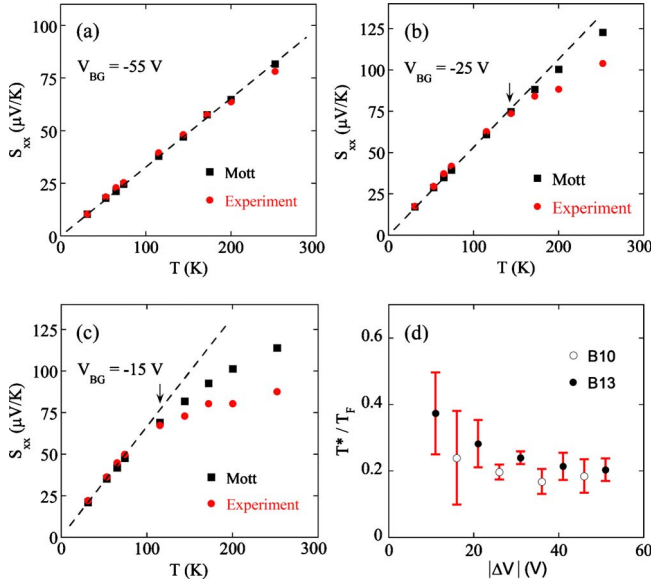

FIG. 3. (Color online) [(a)–(c)] Temperature dependence of the TEP, measured and calculated from the Mott relation, at (a) $V_{BG}=-55$ V, (b) $V_{BG}=-25$ V, and (c) $V_{BG}=-15$ V. Dashed line is a guide to eyes and the solid arrow indicates $T^*$, at which the measured TEP starts deviating from the Mott relation. (d) The ratio of $T^*$ to $T_F$ for two different samples.

As seen in Fig. 3(a), for $V_{BG}$ sufficiently away from the CNP, $S_{xx}^{meas}$ well follows $S_{xx}^{Mott}$ with a linear $T$ dependence. It indicates a weak electron-phonon interaction in bilayer graphene as a phonon-drag component generates the nonlinear $T$ dependence.[5] The linear $T$ dependence of the TEP in both bilayer and monolayer graphene[1–3] is in sharp contrast with that in graphite,[28] where a large negative peak appears at low temperatures due to the dominant phonon-drag contribution. Thus, it will be interesting to study how the phonon-drag effect evolves from monolayer graphene to graphite as the number of graphene layer increases.

As $V_{BG}$ approaches the CNP, $S_{xx}^{meas}$ deviates from $S_{xx}^{Mott}$ at a certain onset temperature $T^*$, around which both $S_{xx}^{meas}$ and $S_{xx}^{Mott}$ also deviate from the linear $T$ dependence [Figs. 3(b) and 3(c)]. Moreover, as indicated by arrows in the figures, $T^*$ decreases with reducing the carrier density. This feature is in clear contrast with that of the monolayer graphene where the linear $T$ dependence holds for $T$ up to room temperature even for the doping levels close to the peak position of TEP.[3]

In the case of $S_{xx}^{Mott}$, the nonlinear $T$ dependence stems from the change in the conductance with $T$ near the CNP [Fig. 1(a)] by the thermally excited carriers.[20,21] On the other hand, to understand the deviation of $S_{xx}^{meas}$ from the Mott relation and the $V_{BG}$ dependence of $T^*$, one needs to recall that the Mott relation is derived from the Sommerfeld expansion, which is valid only for low $T$ ($\ll T_F$). Due to the finite DOS in bilayer graphene at the CNP, $T_F$ of the system is almost an order of magnitude smaller than that of the monolayer graphene for a given charge-carrier density.[29] Thus, as observed in this study, the Mott relation is expected to fail in bilayer graphene at lower temperatures than in monolayer graphene.

We take $T^*$ to be the temperature at which the difference between $S_{xx}^{Mott}$ and $S_{xx}^{meas}$ becomes $3\mu$V/K. As $T^*$ is expected to vary in proportion to $T_F$, the reduced onset temperatures $T^*/T_F$ for two different devices are plotted in Fig. 3(d) as a function of $|\Delta V|=|V_{BG}-V_{CNP}|$. For most of $V_{BG}$'s, $T^*/T_F$ becomes $\sim 0.2$, which is comparable to the theoretically predicted value for the monolayer graphene.[5] Near the CNP, however, $T^*/T_F$ appears to show a slow increase. Although this possible upswing of $T^*/T_F$ potentially leads to new physics, its character should be further clarified with cleaner graphene by overcoming the effect of electron-hole puddles.[23]

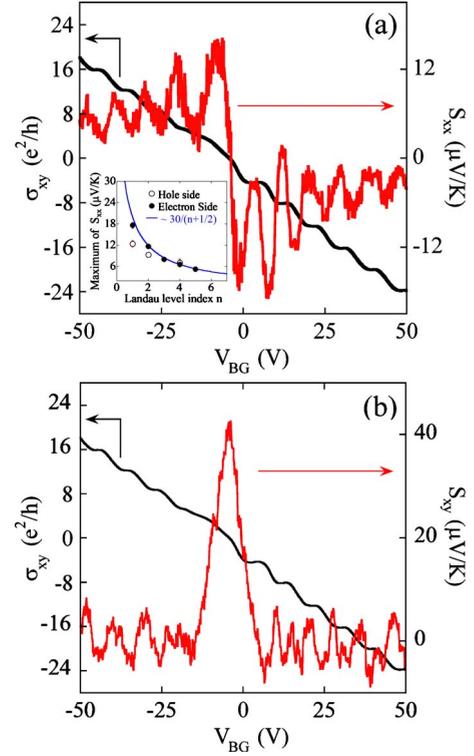

FIG. 4. (Color online) Hall conductivity $\sigma_{xy}$ (solid line) together with (a) $S_{xx}$ and (b) $S_{xy}$ as a function of $V_{BG}$ at temperature of $T=15$ K and in magnetic field of 6.9 T. Inset: peak values of $S_{xx}$ as a function of the Landau-level index $n$, in both hole (open circle) and electron (filled circle) sides. Solid line is $1/(n+1/2)$ fits to the peak values.

In Figs. 3(b) and 3(c), $S_{xx}^{meas}$ also shows a saturating tendency for high values of $T/T_F$. The saturation of TEP was theoretically discussed in monolayer graphene by considering different scatterings[5] due to short-range disorder, acoustic phonons, and charged impurities. As no corresponding theoretical analysis exists for bilayer graphene our results will be useful in investigating controversial scattering mechanisms in bilayer graphene.[11–14]

In a perpendicular magnetic field, $V_{BG}$ dependence of $S_{xx}$ and $S_{xy}$, and the Hall conductivity $\sigma_{xy}$ were measured [see Figs. 4(a) and 4(b)]. $\sigma_{xy}$ exhibits a series of plateaus at $\pm 4Ne^2/h$ for integer $N \geq 1$, which is from the eightfold degeneracy of the zero-energy Landau level.[30] As shown in Figs. 4(a) and 4(b), both $S_{xx}$ and $S_{xy}$ oscillate with $V_{BG}$, similar to the feature observed in a two-dimensional electron gas (2DEG) (Ref. 31) and in monolayer graphene[1–3] in the quantum Hall regime. The oscillations emerge as completely filled Landau levels do not carry entropy, resulting in vanishing $S_{xx}$ and $S_{xy}$ for the chemical potential $\mu$ located between the Landau levels.

When $\mu$ lies at the center of a Landau level, however, $S_{xx}$ becomes maximal. The peak values of $S_{xx}$ are predicted to have a series of fractional values as a function of the Landau level index $n$; $S_{xx}^{peak}=(k_B/e)\ln 2/\Sigma_n f(E_n)$, where $f(E)$ is the Fermi distribution function and $E_n$ is the energy of the $n$th Landau level.[32] $S_{xx}^{peak}$ becomes $\ln 2(k_B/e)/(n+1/2)$ for a 2DEG (Ref. 32) and $\ln 2(k_B/e)/n$ for monolayer graphene.[3] For bilayer graphene, since the $n=0$ Landau level is twice more degenerate than the other levels,[30] $S_{xx}^{peak}$ becomes identical to that of a 2DEG. The peak values of $S_{xx}$ shown in the inset of Fig. 4(a), for both hole and electron sides, roughly follow the $1/(n+1/2)$ dependence. However, the maximum values of $S_{xx}$ are about one half of the expectation, presumably due to the thermal broadening of the Landau levels together with the presence of disorder.[3,31] The Landau-level broadening can be different for each level, which may be the reason for the deviation of $S_{xx}^{peak}$ at $n=1,2$ in the hole side from the $1/(n+1/2)$ fitting.

Finally, let us focus on the Nernst signal, $S_{xy}$. In measurements on the Nernst effect in graphite, a three-dimensional system, the band dispersion along the $c$ axis was found to lead to a maximum of $S_{xy}$ at the center of a Landau level.[15] However, as in a 2DEG (Ref. 31) as well as in monolayer graphene,[1–3] $S_{xy}$ in Fig. 4(b) vanishes at the center of Landau levels, which indicates that a stack of graphene consisting of up to two layers remains to be two dimensional.

## IV. SUMMARY

We measured the thermoelectric power in bilayer graphene at various temperatures and charge-carrier densities. The low-$T$ results of TEP well follows the Mott relation with the hyperbolic band structure of bilayer graphene. The Mott relation of TEP holds well with a linear $T$ dependence for low values of $T/T_F$, signifying negligible phonon-drag effect in bilayer graphene. A deviation starts taking place for high $T/T_F$ along with a saturating tendency of the TEP. These features are not observed in monolayer graphene because its $T_F$ is almost an order of magnitude higher than that of a bilayer. Since the crossover temperature in bilayer graphene is well within the experimentally accessible range, behavior of the TEP in a high-$T$ limit can be conveniently investigated. In a high magnetic field, we also observed oscillatory TEP and the Nernst signals, which are explained by the eightfold degeneracy of the zero-energy Landau level and the two dimensionality of the system, which yields the results distinctive from monolayer graphene as well as graphite.

## ACKNOWLEDGMENTS

Authors thank S. Kettemann and A. Hinz for private communications on the phonon-drag effect in graphene and critical reading of the manuscript. We also thank S. Das Sarma, E. H. Hwang, and E. Rossi for valuable discussion. This work was supported by the National Research Foundation of Korea through Acceleration Research Grants No. R17-2008-007-01001-0 and No. 2009-0083380.